\documentclass{aa}  
\usepackage{xcolor}

\usepackage{graphicx}
\usepackage{txfonts}

\begin{document}

   \title{Excess of substructure due to primordial black holes }

   \author{P. E. Colazo.\fnmsep\thanks{E-mail: patricio.colazo@mi.unc.edu.ar}
          \inst{1,2}
          \and
N. Padilla
          \inst{2,3}
          \and
F. Stasyszyn
          \inst{2,3}
           }

   \institute{Facultad de Matemática, Astronomía, Física y Computación, UNC, Argentina\\
         \and
             Instituto de Astronomía Teórica y Experimental, CONICET--UNC, Argentina\\
         \and
             Observatorio Astronómico de Córdoba, UNC, Argentina\\
             }
   \date{Received 19 May 2025 / Accepted 31 July 2025}

  \abstract
{In this paper we explore the impact of primordial black holes (PBHs) on the abundance of low mass haloes and subhaloes in the dark and low stellar mass regime, and examine how these effects can be measured through fluctuations in strong lensing and brightness fluctuations in clusters of galaxies, providing potential ways to constrain the fraction of dark matter in PBHs.} 
{Various dark matter candidates leave unique imprints on the low mass range of the halo mass function that can be challenging to detect. Among these are the hot and warm dark matter models that predict a reduced abundance of low mass structures compared to the cold dark matter with a cosmological constant ($\Lambda$CDM) model.  Models with PBHs also affect this mass range, but in the opposite direction, producing an increase in these low mass objects. By examining lensing perturbations in galaxy clusters, constraints can be placed on the low mass subhalo abundance and, therefore, on these different models for dark matter. We aim to provide predictions useful for this type of perturbations for the PBH case. Additionally, we examine the abundance of haloes and subhaloes in the range where the stellar mass to halo mass relation is steeply increases, which could be compared to brightness fluctuations in clusters of galaxies due to low mass satellites with low luminosities.}
{We ran cosmological simulations using the {\small SWIFT} code, comparing a fiducial model with alternative inflationary models both with and without PBHs.}
{We find a significant excess of substructure in the presence of PBHs compared to the $\Lambda$CDM model, without altering the abundance of high mass haloes at redshift zero. This increase is of up to a factor of six for extended PBH mass functions with an exponential cut-off at $M_{\rm PBH}=10^2M_\odot$ in the range of parameter space where they could make up all of the dark matter.  Similar increases are also produced when this fraction is smaller, even at sub-percent levels, for PBHs that show an exponential cut-off in their mass function at masses $M_{\rm PBH}=10^4M_\odot$.}
{}
\titlerunning{Excess of substructure }

   \maketitle


\section{Introduction}

Given the diversity of dark matter (DM) candidates, constraining them through observables is essential to resolving the nature of DM. Primordial black holes (PBHs), among these potential candidates, are hypothesised to have formed in the early Universe through alternative inflationary models \citep{Inomata_2017, Clesse_2015}. 
 
Primordial black holes can be described using two primary scenarios: the fixed conformal time (FCT) or horizon crossing (HC). In the FCT scenarios, all PBHs form nearly simultaneously at a specific time during the early radiation era, typically triggered by global mechanisms such as bubble collisions or phase transitions, which define the formation epoch and mass scale. In contrast, the HC scenarios link PBH formation to the moment density fluctuation re-enters the Hubble horizon: Smaller masses cross earlier and collapse first, while larger PBHs form later. This process introduces a natural mass hierarchy (for more details, see \citet{Sureda_2021}). Depending on the primordial power spectrum, these can give rise to either a monochromatic mass distribution (power spectrum with a spike) or an extended mass function (e.g. power spectrum with blue index). Each of these functions allow for different observational constraints across a range of masses on the fraction of PBHs contributing to the total DM density \citep{Carr_2021, Sureda_2021, Padilla_2021, Auclair_2023, NANOGrav_2021, Carr_2017, Sasaki_2018, Niikura_2019}.

The MACHO project placed strong restrictions over the fraction of PBHs that could exist as DM. This project aimed to find compact objects in the Milky Way \citep{Alcock_machos} but found less than what would be expected if PBHs of roughly one solar mass made up most of the DM (for a comprehensive overview on these constraints, see \citet{Carr_2023}). More recently, with the detection of gravitational waves from black hole mergers by the LIGO-Virgo-KAGRA collaboration \citep{abbott_2023}, the idea about PBHs was reawakened. They are now considered potential explanations for several cosmological phenomena, such as gravitational waves due to PBH mergers \citep{Bird_2016, Sasaki_2016, Raidal_2017}
or a way to produce the baryon asymmetry \citep{Carr_1974, Carr_2023, Liu_2022}. 

These black holes have also been proposed as solutions to several cold dark matter (CDM) tensions, including the core-cusp controversy (see \citealt{Boldrini_2020}) and the Hubble tension \citep{Li_2023, Carr_2020}, and to alleviate the tension due to the abundance of high-redshift quasars (\citep{boyuan_2022_stars, Kashlinsky_2021, Goulding2023, Kokorev2023}) and massive  galaxies \citep{Gouttenoire_2023, Su_2023, Colazo_II} observed recently by James Webb Space Telescope (JWST) observations. Additionally, PBHs could serve as seeds for magnetic fields in the Universe \citep{Araya_2021, Papanikolaou2023PhRvD,Padilla_2024} and contribute to the stochastic gravitational wave background \citep{Agazie_2023, Yi_2023}.

Primordial black holes also modify the halo mass function, enhancing the abundance of low mass haloes compared to CDM, particularly through small-scale power \citep{Inman_2019, Colazo_II, Zhang__2024, Matteri, Liu_2022}. The warm dark matter (WDM) model offers an alternative to CDM and affects the halo mass function in the opposite way, by suppressing the formation of low mass haloes due to the cut-off in the initial power spectrum at lower wavenumbers than CDM.  
Several studies \citep{colin_substructure_2000, zentner_halo_2003, he_extending_2023} have shown that WDM would reduce the substructure compared to CDM. 

Early studies by \citet{pinkney_evaluation_1996} and \citet{kochanek_tests_2004} presented different results on whether the inferred substructure is better fit by the CDM or WDM model. In particular, \citet{dalal_direct_2002} found that $\Lambda$CDM is sufficient to explain strong lensing observations.  More recently,  strong gravitational lensing has proven effective in revealing dark subhaloes \citep{nightingale_2023, Vegetti_2009, li_2016, Enzi_2021}. Other models such as fuzzy dark matter (FDM) offer alternative explanations for the observed substructure \citep{elgamal_no_2024}. These models could explain the abundance of structures more massive than $10^9 M_\odot$ while also providing a core unlike CDM that produces a cusp \citep{Salucci_2019}. However, FDM predicts a suppression of small-scale structure similar to WDM models, which may be in tension with recent observational claims of abundant low mass subhaloes \citep{Banik_2021}.

A common denominator in these studies is substructure, which is essential for telling apart these DM models.  Observations such as fast radio burst (FRB) pulses \citep{xiao_detecting_2024} and quasar magnification bias are also promising methods for detecting substructure.  Alternatively, tidal streams in the stellar halo suggest the existence of low mass subhaloes, as predicted by CDM \citep{weinberg_cold_2015}.

We refer to haloes with masses below $10^9,M_\odot$ as dark haloes throughout this work \citep{Mena_2019, Gow_2020, Villanueva_2021, Ziparo_2022, Byrnes_2024}. Although there is ongoing debate about the regime between $10^8$ and $10^9,M_\odot$, where a fraction of haloes may host ultra-faint dwarf galaxies \citep{Kim_2024}, these objects are expected to be extremely faint and effectively invisible at cosmological distances. However, their gravitational imprint—as subhaloes—should still be detectable through lensing perturbations.

The subhalo mass function has been extensively studied. \citet{gao_subhalo_2004} showed that substructure increases with halo mass and decreases with formation redshift and halo concentration. In CDM, the unevolved mass function is scale-free, while WDM models predict a cut-off at relatively high masses, although possibly within the dark halo mass regime  \citep{he_extending_2023, elahi_subhaloes_2009}.

Hydrodynamic simulations have revealed the importance of baryonic processes in altering the formation and evolution of DM haloes. Studies by \citet{dolag_substructures_2009} and \citet{jia_effect_2020} show a reduction in substructure compared to DM-only simulations, particularly in the low mass regime. Dust production also varies between CDM and WDM, with CDM producing significantly more dust due to its higher substructure \citep{ussing_using_2024}.

Future missions such as Euclid, LSST, JWST, and Roman will improve our understanding of substructure \citep{Jwst,LSST,Euclid}. As an example, the upcoming Nancy Grace Roman Space Telescope \citep{Roman} will use microlensing to detect dark substructure, providing competitive constraints on DM models, including PBHs \citep{pardo2021detecting, Tilburg_2018}. Machine learning approaches are also expected to constrain subhaloes in the $10^9 - 10^{9.5} M_\odot$ range \citep{tsang_substructure_2024} by employing tens of thousands of galaxy-galaxy strong lensing systems anticipated to be found in upcoming surveys by the end of the decade. Strong lensing studies combining arcs and flux ratios will further refine these constraints \citep{gilman_turbocharging_2024}.

In order to provide predictions of substructure excess that can be tested with observations of strong lenses, in this study, we simulate a CDM + PBH scenario to numerically follow the enhancement of small-scale substructure. With these simulations, we will be able to study its possible observational detection via strong lensing and also via brightness fluctuations. This is  done by adopting power spectra of density fluctuations with increased power at small scales beyond the observable range including the Poisson effect due to the discreteness of PBHs. Isocurvature effects \citep{Liu_2022} are not included since these are not expected in the range of PBH mass functions we explore. In addition, we adopted a Press–Schechter mass function to model the PBH population \citep{Sureda_2021}. 
Section~\ref{sec: Initial condition} provides a description of the initial conditions and simulation setup. Results are presented in Section~\ref{sec:discussion}, and conclusions are summarised in Section~\ref{sec:conclusions}.

\begin{table}
\caption{Simulation parameters.}             
\label{tab:simulacion_parametros}
\centering
\renewcommand{\arraystretch}{1} 
\begin{tabular}{@{}c p{3.6cm} c@{}}
    \hline
    Parameter & Description & Value\\
    \hline
    \multicolumn{3}{@{}l}{{Shared simulation parameters}} \\
    \hline
    $N_{\text{DM}}$ & Number of DM particles & $1024^3$ \\
    $\text{soft}_{\text{com}}$ & Softening length & $14.8\ \text{kpc}$ \\
    $z_{\text{init}}$ & Initial redshift & $1200$ \\
    $\Omega_{\text{cdm}}$ & CDM density parameter & $0.267$ \\
    $\Omega_{\text{b}}$ & Baryon density parameter & $0.049$ \\
    $\Omega_{\Lambda}$ & Dark energy parameter & $0.684$ \\
    $\sigma_8$ & RMS at 8 Mpc/h & $0.8118$ \\
    $h$ & Hubble parameter & $0.673$ \\
    \hline
    \multicolumn{3}{@{}l}{{PBH model parameters}} \\
    \hline
    $k_{\text{piv}}$ & Pivot scale & $10\ \rm{cMpc}^{-1}$ \\
    $n_b$ & Blue spectral index & $2.5$ \\
    $f_{\text{PBH}}$ & DM fraction in PBHs & $1$ \\
    $M^*_{\text{PBH}}$ & Characteristic PBH mass & $10^2\ M_\odot$ \\
    \hline
    \multicolumn{3}{@{}l}{{High-resolution simulation}} \\
    \hline
    $L_{\text{box}}$ & Box size & $20\ \rm{cMpc}/h$ \\
    $M_{\mathrm{mp}}$ & Particle mass & $10^{5.986}\ M_\odot$ \\
    \hline
    \multicolumn{3}{@{}l}{{Low resolution simulation}} \\
    \hline
    $L_{\text{box}}$ & Box size & $35\ \rm{cMpc}/h$ \\
    $M_{\mathrm{mp}}$ & Particle mass & $10^{6.716}\ M_\odot$ \\
    \hline
    \multicolumn{3}{@{}l}{{Simulation from previous work\tablefootmark{a}}} \\
    \hline
    $L_{\text{box}}$ & Box size & $205\ \rm{cMpc}/h$ \\
    $M_{\mathrm{mp}}$ & Particle mass & $10^{9.019}\ M_\odot$ \\
    $z_{\text{init}}$ & Initial redshift & $127$ \\
    $f_{\text{PBH}}$ & DM fraction in PBHs & $0.005$ \\
    $M^*_{\text{PBH}}$ & Characteristic PBH mass & $10^4\ M_\odot$ \\
    \hline
\end{tabular}
\tablefoot{
Main set of simulation parameters, including the cosmological parameter values. 
For convergence testing, three additional simulations were run (high resolution), 
differing only in box size, which was set to $20\,\mathrm{Mpc}/h$.\\
\tablefoottext{a}{For completitude, we included the parameters of the Simulation parameters from \citet{Colazo_II}.}
}
\end{table}

\section{Simulations}

In this section, we describe the setup and execution of the simulations used in this study. We start by describing the power spectra used to construct the initial conditions in Subsection \ref{sec: pdk}, then we outline the careful design of the initial conditions for our runs in Subsection \ref{sec: Initial condition}, and we list the set of simulations run for this work in Subsection \ref{sec: sets}.

\subsection{Power spectrum}
\label{sec: pdk}

We adopted the following primordial power spectrum,
\begin{equation}
P_{\mathrm{primordial}}(k) = A_S \left( \frac{k}{k_0} \right)^{n_s},
\end{equation}
where $A_S$ is the typical amplitude of fluctuations at the pivot scale $k_0 = 0.05\, \mathrm{Mpc}^{-1}$, and $n_s$ is the spectral index that controls the tilt of the power spectrum. When $n_s < 1$, it is known as a red index; when $n_s > 1$, it is referred to as a blue index. In standard CDM models, $n_s$ is slightly below unity, $n_s = 0.965 \pm 0.004$ \citep{Planck_2020}. This spectrum has been evolving since the end of the inflationary era, with changes encoded in the transfer function, and can be measured using observations of the cosmic microwave background (CMB).

In our simulation of PBHs, the power spectrum follows the $\Lambda$CDM model up to a scale $k_{piv}$, well within the non-linear regime where the power spectrum shape is not yet constrained. To construct this power spectrum, we set $k_{\mathrm{piv}} \approx 10\, \mathrm{Mpc}^{-1}$, consistent with the approach of \citet{Sureda_2021}. Introducing PBHs requires an inflationary model with enhanced primordial power on small scales \citep{Inomata_2017, Clesse_2015, Kawasaki_2013, Gupta_2020}. This increase occurs at $k > k_{\mathrm{piv}}$, where the new spectral index is denoted by $n_b$ (blue index), with $\varepsilon$ serving as a normalisation factor,
\begin{equation}
\Tilde{P}_{\mathrm{primordial}}(k) = 
\begin{cases} 
A_S \left( \frac{k}{k_0} \right)^{n_s}, & \text{if } k < k_{\mathrm{piv}}, \\ 
A_S \varepsilon \left( \frac{k}{k_0} \right)^{n_b}, & \text{if } k > k_{\mathrm{piv}}.
\end{cases}
\end{equation}
\label{equ:broken_power}

The blue spectral index, $n_{\mathrm{b}}$, affects the steepness of the exponential decline in the PBH mass function. Since a primordial power spectrum with a break to a blue index does not necessarily  imply the formation of PBHs, we also explore a simulation with the blue index but without Poisson noise from PBHs ('NB'), to explore this possibility. Ruling out a blue index automatically rules out a wide set of models for extended PBH mass functions.

When PBHs are present, their existence as discrete, massive particles introduces a significant Poisson effect on the gravitational potential, modifying the evolution of fluctuations \citep{Padilla_2021}. Therefore, the power spectrum includes this effect, and we account for the fraction of DM composed of PBHs, ($f_{\mathrm{PBH}}$).

\begin{equation}
P(k,z) = \Tilde{P}_{\mathrm{primordial}}(k) \, T^2(k) \, D_1^2(z) + f_{\mathrm{PBH}}^2 \, P_{\mathrm{Poisson}}^{\mathrm{PBH}}(k,z),
\end{equation}
where $T(k)$ is the transfer function and $D_1(z)$ is the linear growth factor. We fixed $f_{\mathrm{PBH}} = 1$, although the resulting spectrum is very similar (differences are of less than $10\%$) to a model used in \citet{Colazo_II} for which the characteristic mass is higher and $f_{\mathrm{PBH}} <<1$.  We note that $P_{\mathrm{Poisson}}^{\mathrm{PBH}}(k,z)$ evolves with both redshift and $k$-space (see \citealt{Padilla_2021} for more details).
 
The parameter $f_{\mathrm{PBH}}$  directly determines the number density of PBHs. We selected the maximum value allowed under current constraints for an FCT (fixed conformal time) mass function \citep{Sureda_2021}, setting $f_{\mathrm{PBH}} = 1$ for a characteristic PBH mass of $100\, M_\odot$. Notice that this fraction is expected to be lower if taking into account gravitational wave constraints, but these remain under debate. It is worth noting that $\mu$-distortion analyses and CMB constraints from \citet{Gow_2020} and \citet{Byrnes_2024}, as well as additional limits from the 21\,cm signal \citep{Mena_2019, Villanueva_2021} and the X-ray background \citep{Ziparo_2022}, tend to disfavour $f_{\mathrm{PBH}} = 1$ for $M_{\mathrm{PBH}} \simeq 100\, M_\odot$, but many of these studies use either monochromatic or narrow log-normal mass distributions, which yield strong constraints due to their sharp mass localisation. However, as shown by \citet{Sureda_2021}, in extended mass functions such as those derived from Press–Schechter theory and smooth primordial power spectra with a blue index, the constraints can be significantly relaxed. This is because only a small fraction of the total PBH density lies near $M^\ast$, with the majority spread across other mass scales.

We selected the maximum value allowed under current constraints for an FCT mass function \citep{Sureda_2021}, setting $f_{\mathrm{PBH}}=1$ for a characteristic PBH mass of $100 \ M_\odot$, disregarding gravitational wave constraints, which remain under debate. The Horizon-crossing mass function, distinct from FCT, generally produces a steeper profile with the same blue index, yielding a less pronounced effect.

\begin{figure}
    \centering
    \includegraphics[width=0.5\textwidth]{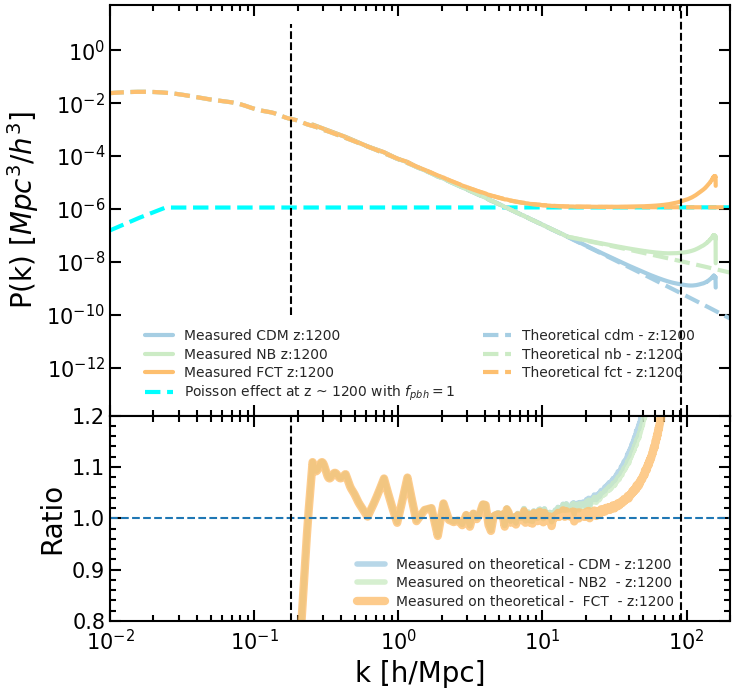}
    \caption{Complete power spectrum at $z=1200$ for the initial condition.  Theoretical linear predictions and measured power spectra of initial conditions estimated using Pylians3 are shown in the top panel. The blue line shows the initial spectrum for the $\Lambda CDM$ model. The initial power spectrum with all contributions from PBHs is shown in orange. The  case with NB but no Poisson contribution (no PBHs) is shown in green. The cyan line shows the contribution due to the Poisson effect. The vertical dash-dotted grey lines show the location of $k_{\text{min}}=\frac{2\pi}{L_{box}}$ and $k_{\text{nyq}}$. Bottom: Ratio between the measured power spectra against the linear theoretical model.
    }
    \label{fig:ICS_med}
\end{figure}

In Figure~\ref{fig:ICS_med}, we summarise the complete power spectrum both from theory and measured from the particle distribution of the initial conditions at $z=1200$. For the analysis, we use Pylians3 \citep{Pylians} on initial conditions generated by {\small MUSIC2-MONOPHONIC} (\citealt{MUSIC2-MonophonIC}, see the simulation details in S\ref{sec: pdk} and in Table \ref{tab:simulacion_parametros}). For all cases, solid lines show the measured spectra, while dashed lines represent the theoretical linear estimation from CAMB \citep{Camb}. The initial $\Lambda$CDM power spectrum without PBHs at $z = 1200$ is shown in blue. The orange line represents the full power spectrum, with contributions from the primordial spectrum, the blue index in green, and the Poisson effect in cyan. The lower panel shows the ratio of the measured power spectra to the theoretical expectation, highlighting aliasing effects at small scales. On large scales, all three simulations align to the same values.

\subsection{Initial conditions}
\label{sec: Initial condition}

In cosmological simulations, initial conditions are generally considered to be pseudo-linear, characterised by Gaussian distributions of fluctuations. It is essential to confirm that these conditions are met. Difficulties arise when one adopts increased power at smaller scales which can produce fluctuations that do not fall within the quasi linear regime required for initial conditions. 

Three primary criteria must be satisfied for our simulations.

\begin{figure}
    \centering
    \includegraphics[width=0.5\textwidth]{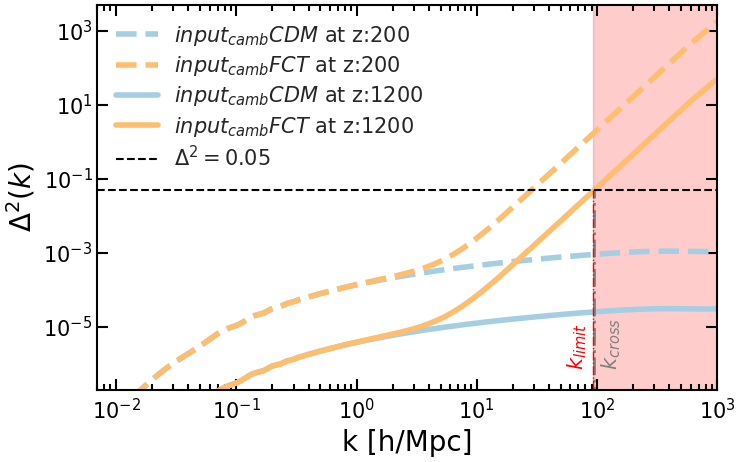}
    \caption{Illustration of the factors influencing the generation of initial conditions in our simulations. The plot shows  $\Delta^2$ as a function of wavenumber, highlighting the upper limit of $\Delta^2 < 0.05$ for appropriate 3LPT application. The red-shaded area indicates the resolution limit for $k_{\mathrm{cross}}$ ($94.78$, marked by the grey dotted line). If the threshold falls within this region, the simulation can begin at the corresponding redshift. Power spectra are displayed for two simulations (blue and orange) at different redshifts (solid vs. dashed lines), identifying the earliest redshift where $k_{\mathrm{cross}} > k_{\mathrm{limit}}$. This criterion determines the starting redshift. For the FCT model, this condition applies at $z \sim 1200$.
}
    \label{fig:2LPT_conditions}
\end{figure}

\begin{enumerate}
    \item {Wavenumber limit (${k_{\mathrm{limit}}}$)}: This parameter depends on the simulation configuration, which involves carefully selecting three critical factors:
    \begin{enumerate}
        \item The minimum halo mass to be resolved.
        \item The total number of particles in the simulation.
        \item The minimum particle count required to identify individual haloes.
    \end{enumerate}
    
    These factors determine the simulation box size needed. Our objective is to evaluate the abundance of subhaloes relative to the CDM scenario, particularly considering the presence of primordial black holes. We adopted a minimum threshold of 30 DM particles per halo. To ensure the robustness of our analysis, we check increasing the threshold to 100 (see Appendix~\ref{sec:convergence}) with minimum changes in the results. Therefore, for a box size of 35 Mpc$/$h we adopted a total particle count of $1024^3$ in our simulations.\footnote{We conducted three additional simulations with smaller box sizes of $20\, \text{Mpc}/h$, resolving haloes around $\sim 8.0 \times 10^8\, M_\odot$ with at least 100 particles. The results exhibit excellent agreement.}
    
    An iterative algorithm was developed to optimise the selection of parameters, enabling us to meet the resolution requirements while maximizing the simulated volume. The wavenumber for these resolutions is  $k_{\mathrm{limit}} = 91.9\, h/\mathrm{Mpc}$  \footnote{$k_{\mathrm{limit}} =  160.8\, h/\mathrm{Mpc}$ in case of simulations with box sizes of $20\, \text{Mpc}/h$} calculated as \( k_{\mathrm{limit}} = {2 \pi}/{L_{\mathrm{box}}} \cdot {N_{\mathrm{DM}}^{1/3}}/{2} \)  
    (we note that $k_{\mathrm{nyq}}= {2 \pi}/{L_{\mathrm{box}}} \cdot {\mathrm{Grid size}}/{2}$ where, in our case, $\mathrm{Grid size} = 1024 $ was used to compute the power spectrum, i.e. $k_{\mathrm{nyq}} = k_{\mathrm{limit}}$). 
   
    For a halo, the corresponding wavenumber is defined as $k_{\mathrm{min-halo}} = k_{\mathrm{limit}} / 2$. This criterion is crucial due to its proximity to the resolution limit, where aliasing effects may contaminate the power. Aliasing grows significantly near $k_{\mathrm{Nyq}}/2$ (R. Scoccimarro, private communication). Additionally, the Poisson contribution to the power spectrum, which dominates the PBH signal at small scales alters the aliasing pattern by injecting additional noise, effectively acting as a smoothing or 'erasing' mechanism over the contamination.
    Therefore, given our available computing power we prioritise reducing the box size rather than compromising the signal-to-noise ratio within the target wavenumber range ($k_{\mathrm{min-halo}}$).
    
    \item {Small fluctuations}: Generating initial conditions involves modelling the power spectrum using third-order Lagrangian perturbation theory (3LPT, \citealt{MUSIC2-MonophonIC}). This method is effective when
    $\Delta= k^3P(k)/2\pi$, which quantifies the contribution to the variance of matter fluctuations per logarithmic interval in $k$-space, satisfies $\Delta^2 < 0.05$.
    The fundamental requirement for our initial conditions is that the crossing wavenumber,
defined by $\Delta^{2}\!\bigl(k_{\mathrm{cross}}\bigr)=0.05$,
satisfies $k_{\mathrm{cross}}>k_{\mathrm{limit}}$.
 We thus select $z = 1200$ as the starting redshift, with $k_{\mathrm{cross}} = 94.78$ in the FCT case.  We choose to start all simulations from this initial redshift.
     Figure \ref{fig:2LPT_conditions} illustrates $\Delta^2$ as a function of wavenumber, with an upper limit of $\Delta^2 < 0.05$ within the quasi-linear regime. In particular, the FCT model cannot use a starting redshift of $z = 200$ (see the dashed orange line); instead, its initial resolution requirement is met only at a higher redshift, around $z \sim 1200$ (solid). Radiation effects on perturbations are neglected in all simulations, as this work focuses on low redshift regimes.  This omission is present in all models, which makes it possible to compare results across models. 
    
    After taking into account these considerations, we use the {\sc{ \small MUSIC2-MONOPHONIC}} code \citep{MUSIC2-MonophonIC} to generate our initial particle distributions.

    \item  {Consistency with the fiducial model on large scales}: After generating the initial conditions, the  power spectra are re-measured to verify the fidelity of the 3LPT method and its alignment with the theorical models used (see Fig \ref{fig:ICS_med}). Once initial consistency is established, the simulations are ready for execution.

    Numerical errors introduced at this early stage beyond those that checked to be absent via this power spectra comparison are not a major concern, as all simulations share the same random seeds.

\end{enumerate}

\subsection{Simulation sets}
\label{sec: sets}

We conducted three main simulations with sufficient resolution for resolving substructures while maintaining a sample size large enough to capture haloes with sufficiently high masses that could be targeted for strong lensing studies. All simulations start at a redshift of $z = 1200$, the lowest feasible redshift for applying the initial condition for the FCT model described previously; this model has the highest fluctuations on small scales of the three we study here. One simulation follows the standard $\Lambda$CDM cosmology, while the second incorporates a modified blue spectral index. The third simulation includes this blue index but also primordial black holes (PBHs) with a characteristic mass of $M^* = 10^2$ within a Press-Schechter mass function based on the fixed conformal time (FCT) framework \citep{Sureda_2021}. We note that PBHs are not included as individual particles in the simulations; their effects are encoded solely through modifications to the initial power spectrum. These simulations facilitate comparative analysis to assess the impact of two main effects: enhanced small-scale power from alternative inflationary models and the additional gravitational potential contributed by discrete PBHs. The primary properties of these simulations are summarised in Table \ref{tab:simulacion_parametros}.

All simulations were conducted using the {\small SWIFT} simulation code \citep{swift}. We adopted the fiducial cosmological parameters provided by  \cite{Planck_2020}.

Additionally, we ran three supplementary simulations with a box size of $20\, \text{Mpc}/h$ to evaluate the resolution effects on smaller haloes with increased particle counts. These models show excellent agreement with the larger boxes, as shown in the results presented below.

{
Our simulation with PBHs is, for all practical purposes, a proxy for the one used in \citet{Colazo_II}; we also present the parameters of this simulation in Table \ref{tab:simulacion_parametros}. The initial power spectra for these two FCT simulations differ by less than $10\%$, which confirms that the conclusions for the FCT mass function adopted here extend also for the FCT model with higher PBH characteristic mass and lower $f_{\mathrm{PBH}}=0.005$ of \citet{Colazo_II}. Even though our FCT model does not include the isocurvature contribution (because PBHs are not of large enough mass), the large fraction of DM in PBH makes the Poisson contribution similar to the Isocurvature of the more massive and $200$ times lower density PBH population of \citet{Colazo_II}. The resulting effects on structure formation are therefore qualitatively similar, although they arise from distinct black hole distributions and abundances. This suggests that PBH configurations can be degenerate in offering solutions to cosmological tensions as is the case of our earlier work. }

\section{Results and discussion}
\label{sec:discussion}

In this section we analyse the abundance of haloes and subhaloes in our simulations, and make an analysis of the substructure of the most massive haloes which correspond to galaxy groups. We do this to have predictions for possible observables that could tell apart particle DM from PBHs.

We first concentrate on DM haloes. We use the {\small ROCKSTAR} code which uses phase-space information to find all gravitationally bound structures, including main haloes and subhaloes. We ensure that all simulations use the same force resolution to maintain consistent numerical accuracy across the models. As expected, we confirm a higher abundance of dark haloes (low mass haloes) in the NB and FCT models compared to the CDM simulation. To specifically identify subhaloes, we subsequently ran the {\small find\_parent} routine provided with {\small ROCKSTAR}, which determines whether a given halo resides within a more massive host halo, classifying it as a subhalo.

\begin{figure}[t]
    \centering
    \includegraphics[width=0.5\textwidth]{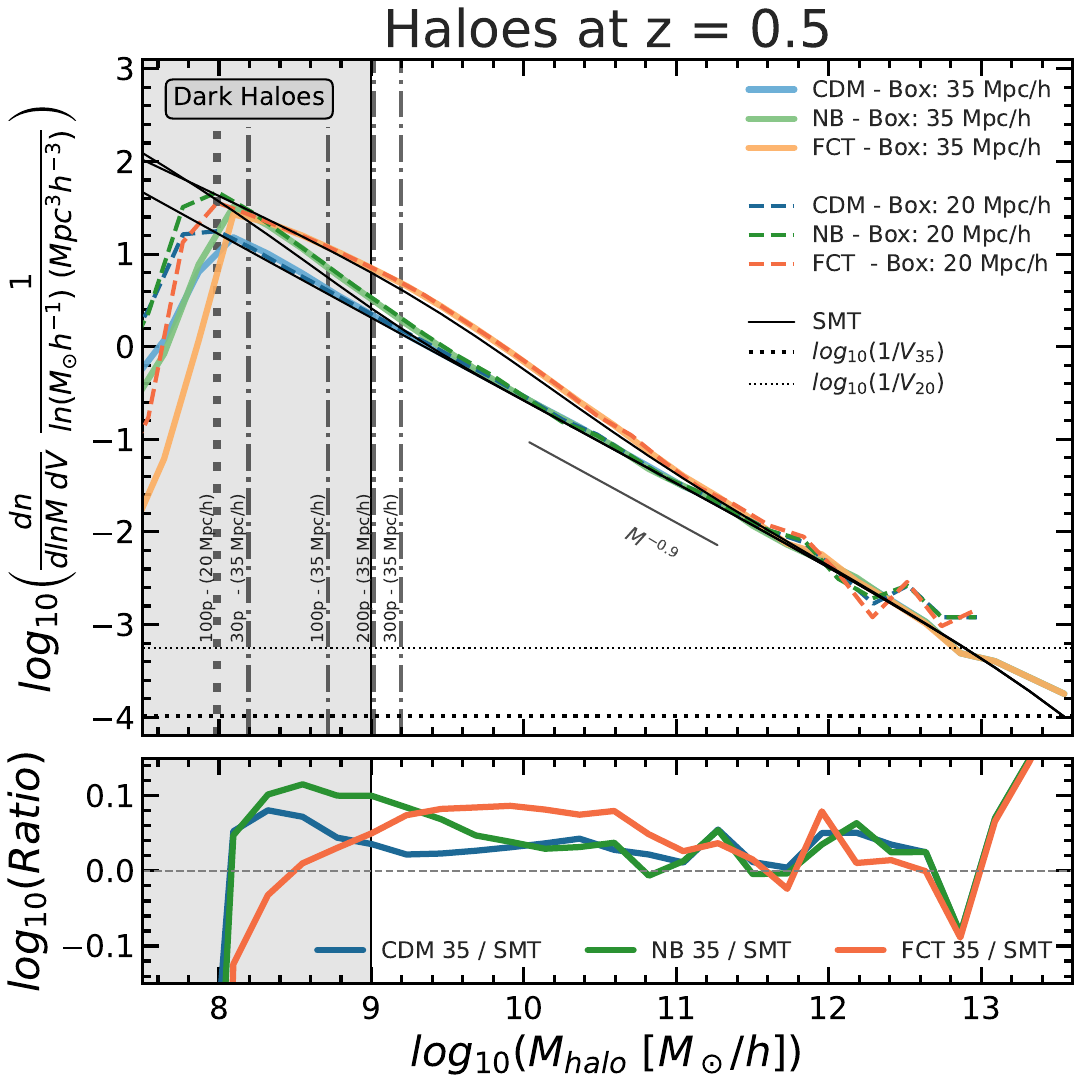}
    \caption{Halo mass functions at $z = 0.5$ for CDM (blue), NB (green), and FCT (orange) simulations. Solid colored lines show haloes with at least 30 particles in the $35\, \mathrm{Mpc}/h$ boxes; dashed lines show higher-resolution runs in $20\, \mathrm{Mpc}/h$ boxes, requiring at least 100 particles per halo. The agreement at low masses is excellent; at high masses, the number of haloes drops more precipitously in the small box, increasing scatter. A power-law slope to the CDM mass function is shown in black $\propto M^{-0.9}$. The grey-shaded area marks the 'dark sector' (haloes without luminous components). Sheth, Mo \& Tormen mass functions (SMT) are included for all models as solid black lines that match quite closely the simulation measurements. The lower panel shows the ratio between the simulation mass functions and SMT in the $35\, \mathrm{Mpc}/h$ boxes, with deviations around or below $10\%$. The horizontal dotted lines indicate the density corresponding to one per volume element for the high- and low resolution simulations. }
    \label{fig:f_mass}
\end{figure}

Figure~\ref{fig:f_mass} shows the halo mass functions at $z = 0.5$. The three simulation sets—CDM (blue), NB (green), and FCT (orange)—are compared, with additional simulations using smaller box sizes of $20 \, \text{Mpc}/h$ shown as dashed lines. We observe excellent agreement at low masses between simulations of the same model but with different resolutions. At the high-mass end, the number of objects decreases and the curves begin to diverge due to low number statistics, reflecting the impact of the reduced simulation volume of the smaller simulation. The short black line in the figure represents a power-law fit to the CDM mass function \citep{elahi_subhaloes_2009}, confirming the expected trend. The long black curves correspond to the Sheth–Tormen mass functions (SMT; \citealt{ST}) computed using the linear theory power spectrum for each model. The partial disagreement between SMT and the simulated curves highlights the importance of addressing these questions using numerical simulations rather than purely theoretical approaches. Finally, The grey-shaded area identifies haloes thought to lack luminous components

\begin{figure}[t]
    \centering
    \includegraphics[width=0.5\textwidth]{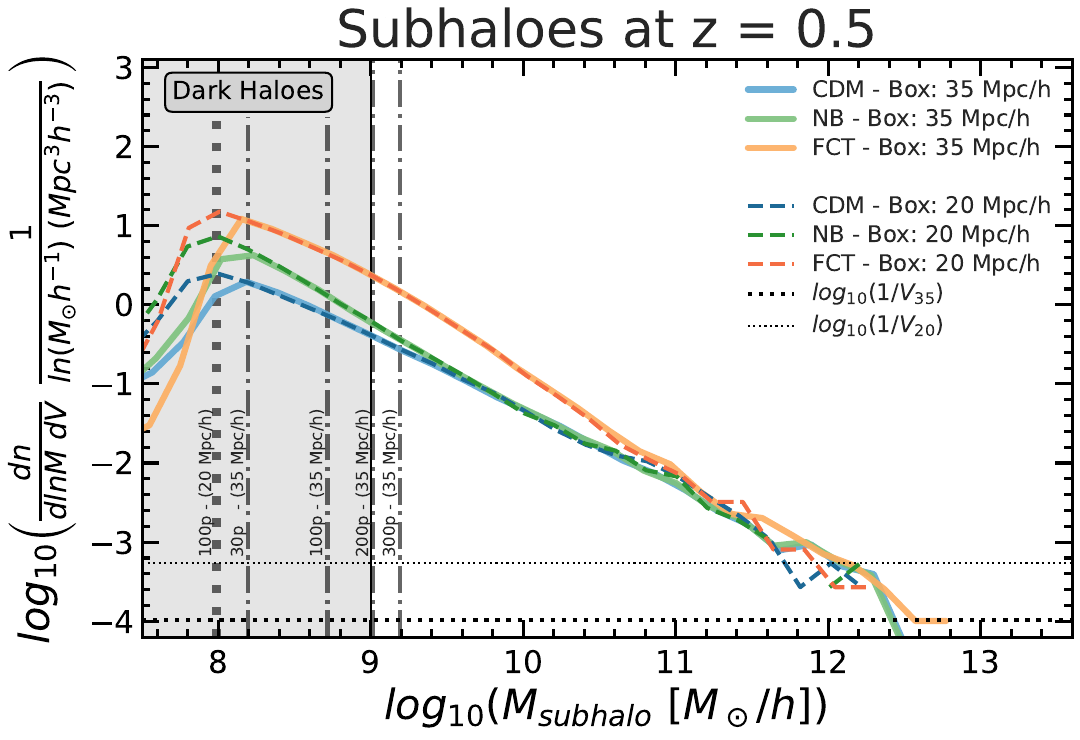}
    \caption{
    Subhalo mass functions at $z = 0.5$ as a function of subhalo mass. Colors and line styles follow the same coding as in Figure~\ref{fig:f_mass}, with solid lines for $35\, \mathrm{Mpc}/h$ simulations and dashed lines for the higher-resolution $20\, \mathrm{Mpc}/h$ runs. Particle number thresholds in the $35\, \mathrm{Mpc}/h$ boxes are varied to illustrate the impact of resolution cuts on the subhalo mass function estimates.
    }
    \label{fig:f_mass_sub}
\end{figure}

An important consistency check is the matching abundance of high mass haloes across all simulation sets. A significant discrepancy would challenge PBH models, as current observations are in reasonable agreement with the abundance of groups and clusters (e.g. \citealt{Abdullah}). 

\begin{figure}[t]
    \centering
    \includegraphics[width=0.5\textwidth]{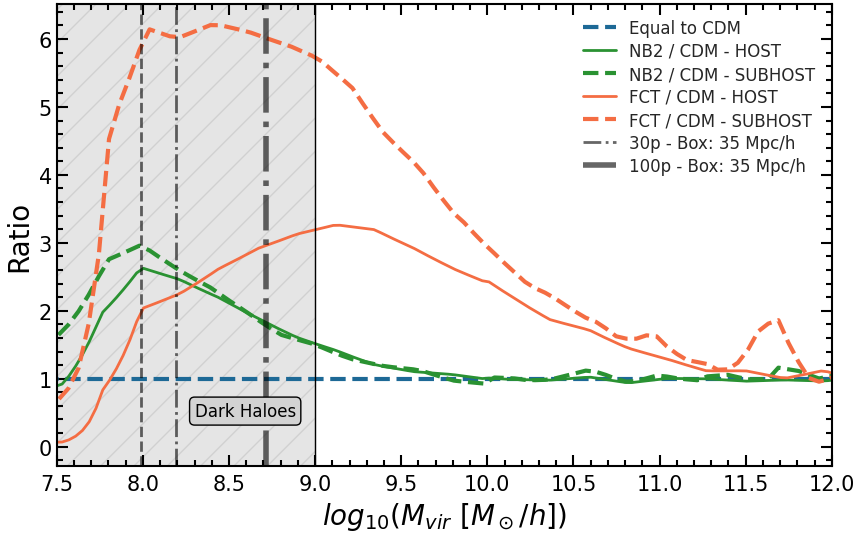}
    \caption{
    Ratio of subhalo (dashed lines) and halo (solid lines) abundances for the NB (green) and FCT (orange) models relative to the CDM baseline, shown as a function of mass. The blue horizontal line marks perfect agreement with CDM. Each curve combines results from simulations with different box sizes (see text for details). A clear excess near $10^8\, M_\odot$ in both NB and FCT models suggests a potentially detectable signal via strong lensing. The grey vertical line indicates the resolution threshold. Notably, subhaloes show larger deviations than haloes, which may enhance detectability in lensing analyses.
    }
    \label{fig:Percentage}
\end{figure}

The models start to diverge at intermediate and low masses, below $10^{11}\, M_\odot$ (see Fig. \ref{fig:Percentage}). We explore two main avenues to detect the effects of PBHs. The first focuses on dark haloes in the mass range $10^8$–$10^9\, M_\odot$, where differences between models are most pronounced. These objects lack luminous components, making them difficult to observe directly, but they can be probed through strong lensing without the need to model galaxy formation \cite{li_2016}. The second approach examines haloes in the next mass regime ($10^9$–$10^{10}\, M_\odot$), where PBHs may impact the abundance of low stellar mass galaxies since this mass range is more likely to contain a significant population of stars. Techniques such as HOD and extended SubHalo Abundance Matching (SHAMe) offer alternative frameworks to link galaxies with haloes (see \citealt{Contreras_2024} for a detailed comparison).

Using Markov Chain Monte Carlo (MCMC) methods and emulators, these approaches generate parameter sets consistent with observed clustering.
This opens an interesting application for our results: PBH-induced modifications to the halo mass function may alter the galaxy-halo connection, potentially affecting the galaxy correlation function.
Although a full exploration of this idea is beyond the scope of this work, future studies could test whether excess halo abundance—particularly at low masses—leaves detectable imprints on the clustering of low stellar mass or low luminosity galaxies through adjusted HOD or SHAM parameters.

Secondly, we look at the abundances of subhaloes. Notice that from this point on, the use of numerical simulations are mandatory, as the complicated physical processes within virialised haloes make it difficult to produce accurate subhalo mass functions analytically.
 Figure ~\ref{fig:f_mass_sub} shows the resulting subhalo mass functions, where it can be seen that the abundance of subhaloes is even more enhanced in the PBH model with respect to $\Lambda$CDM. This figure also shows the results for the smaller boxes ($20\, \text{Mpc}/h$), confirming that even subhaloes in the $10^8$–$10^9\, M_\odot$ range, resolved with 30 particles in the main runs, remain robust. The good agreement across resolutions indicates that numerical noise is not the primary source of the observed substructure excess, which more likely has a physical origin in the enhanced small-scale power in PBH models, which is physically motivated.

To ensure that numerical noise does not dominate our results—particularly spurious subhalo disruption—we tested whether all simulations are equally affected by numerical artefacts.  As shown by \citet{Power}, timestep choice, softening length, particle count, and force accuracy are key factors influencing halo stability. Following the convergence criteria from \citet{van_den_Bosch_2018} and \citet{joop_2019}, we required at least $N_c = 100$ particles per halo and subhalo.

One possible reason for our apparent lack of numerical disruption is discussed by \citet{van_den_Bosch_2018} (see figure \ref{fig:f_mass_sub}). They argue that if this disruption is present, it would minimally affect our comparative analysis, as all simulations share the same softening length and similar numerical parameters. 

In any case, any additional disruption induced by PBH‑enhanced tidal fields would only reinforce our conclusions (that follow from PBHs seeding more early structure than CDM, which leads to a denser environment and thus stronger tidal torques on haloes, \citealt{Pato_2019}), making the observed substructure excess a conservative lower limit, although higher concentrations—due to earlier PBH‑driven halo formation—could partially counteract this effect. A quantitative discussion of numerical effects is presented in Appendix \ref{sec:convergence}.

Figure~\ref{fig:Percentage} shows the abundance ratio of haloes (solid lines) and subhaloes (dashed lines) in the NB and FCT models relative to CDM, as a function of mass. To compute these ratios, we combined the mass functions from both resolutions: below $10^{10}\, M_\odot$ we used the $20\,\mathrm{Mpc}/h$ box, and above that threshold the $35\,\mathrm{Mpc}/h$ box. The blue horizontal line marks perfect agreement with the CDM baseline. The NB and FCT models exhibit a pronounced enhancement in subhalo abundance relative to CDM, particularly in the $10^8$–$10^9\, M_\odot$ range, where FCT reaches up to six times the CDM value. This excess becomes measurable as soon as the first bound haloes form and remains significant down to $z=0.5$, or even to $z=0$. At $z\!\sim\!25$ \citet{perezgonzalez_2025} report ultra–luminous sources whose rapid assembly can be reproduced if a tiny fraction, $f_{\rm PBH}\!\sim\!10^{-7}$, of $\sim10^{4}\,M_\odot$ PBHs accelerates early halo growth \citep{matteri_2025}.  The same mechanism underlies the excess we find in our FCT run, which at $z=0.5$ still exceeds both the signal in main haloes compared with CDM and the  expectation from Sheth–Tormen theory. Together with our previous analysis \citep{Colazo_II}, these results indicate that PBHs can boost structure formation from cosmic dawn onward. While this strengthens the prospects for detection, it may also reflect numerical disruption affecting more strongly low mass CDM subhaloes. This effect is discussed in detail in Appendix~\ref{sec:convergence}. At minimum, the halo-level signal should represent a conservative lower bound for the substructure signal. Additional processes—such as dynamical friction or enhanced merger rates—could further amplify this difference. Although our results are not fully corrected for numerical artefacts, they are not dominated by them either, ensuring that the reported excess is a reasonable lower limit.

If a large galaxy cluster exhibits an excess of dark substructure, detecting the signal represented by the green line (NB) is essential to fulfill one of the necessary conditions for the existence of PBHs. The results between this signal and the orange line could be evidence the presence of PBHs (this signal is governed by the Poisson spectrum, which depends on the fraction of PBHs within the DM component). Extending the PBH parameter space beyond the ones studied here (and in \citealt{Colazo_II}) could allow to use this connection to constrain the parameters of these models.

Although this substructure signal is measured across the full simulation, it may vary slightly with environment. In particular, we expect the gap between models to persist in galaxy clusters, where the denser conditions could even enhance detectability. As our volume does not contain cluster-scale haloes, we focus on the most massive systems available—comparable to galaxy groups. The analysis of this subsample is presented in the following subsection.

\label{sec:prediction}
\subsection{Stacking of the most massive haloes for strong lensing search of dark haloes}

We now focus on the most massive haloes in our simulations due to their potential relevance for strong lensing studies. As shown by \citet{Fo_x_2013}, haloes with masses around $10^{13}\ M_{\odot}$ can produce detectable strong lensing signals. Upcoming surveys are expected to provide sufficiently large samples to enable stacking analyses (see \citealt{LSST_Science_Book}, \citealt{mccarty2024stronggravitationallensingupcoming}).

We select a subsample of the 17 most massive haloes matched in the three larger box simulations, with masses in the range $10^{13.0}\ M_{\odot} < M_{h} < 10^{13.6}\ M_{\odot}$. For each halo, we generated three independent projections along the principal axes, yielding a total of 51 distinct views for stacking analysis.

\begin{figure}[t]
    \centering
    \includegraphics[width=0.5\textwidth]{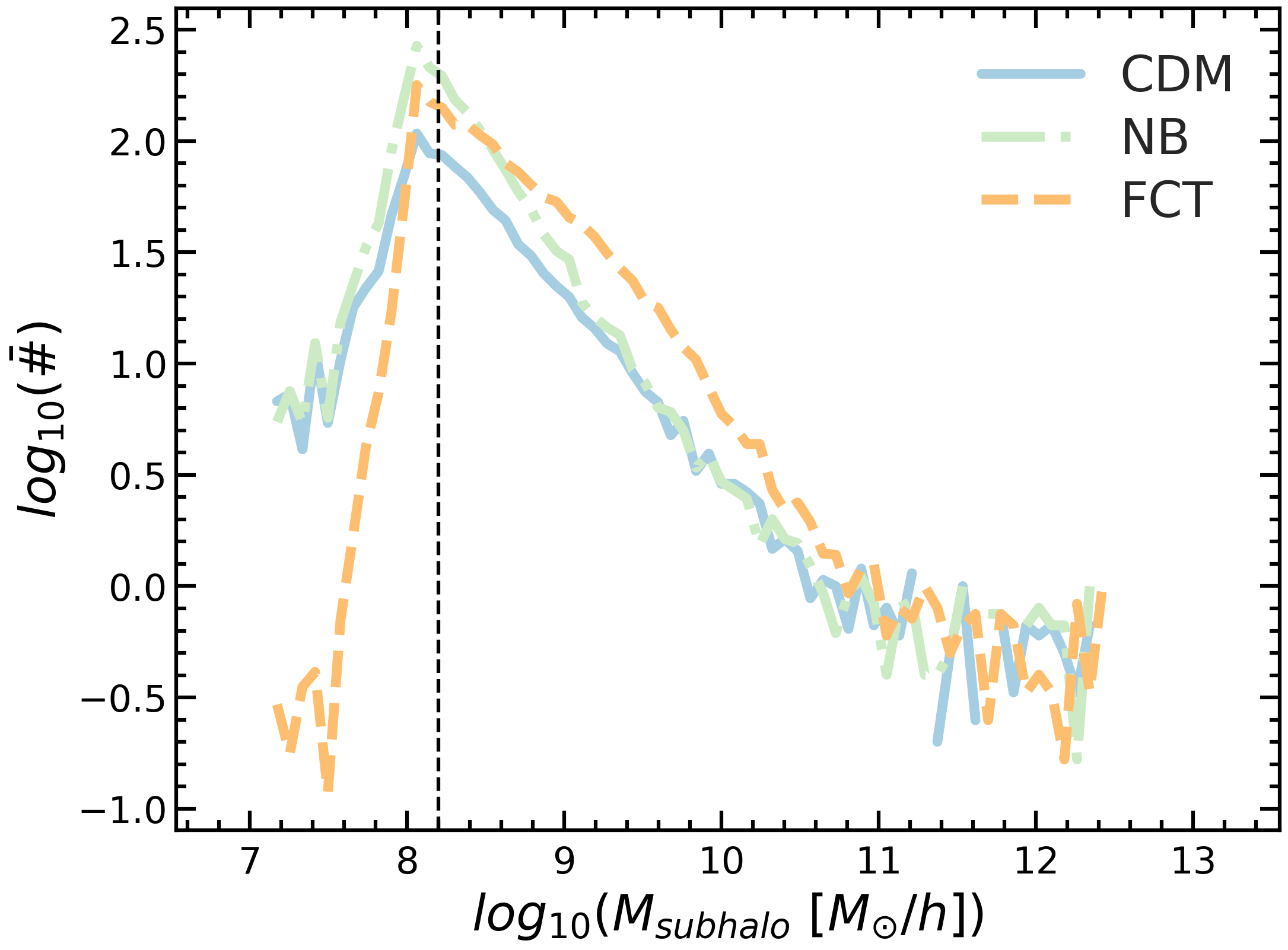}
    \caption{{Mean subhalo mass distribution (with equal arbitrary normalisation across the different models) computed from the 17 most massive haloes in the simulation set. The black dashed line marks the resolution limit adopted throughout the analysis (as defined in Figure~\ref{fig:f_mass}). The three models—CDM (blue), NB (green), and FCT (orange)—show distinct behaviours. The FCT model exhibits a clear excess in the abundance of low mass subhaloes ($M_{vir} < 10^9\ M_{\odot}$), while at intermediate masses, the NB model overtakes the FCT model. This crossover region provides a potential discriminant between scenarios involving enhanced primordial power (NB) and those that also include PBHs (FCT). See text for further discussion.}}
    \label{fig:Funcion_masa_subahalos}
\end{figure}

As a first step, we examine the subhalo mass distribution within the most massive haloes of our simulation sample. Figure~\ref{fig:Funcion_masa_subahalos} displays this distribution, with a lower mass threshold of $M_{h} = 10^{8.2}\ M_{\odot}$ applied consistently, as defined in Figure~\ref{fig:f_mass}. We observe that, close to this resolution limit, the NB model overtakes the FCT model in abundance. The origin of this trend—and its implications for PBH detection—are explored throughout this section.

To control for numerical disruption, we analysed halo concentration distributions. The FCT model shows higher concentrations than CDM and NB at fixed mass, consistent with earlier formation  reported in \citet{Colazo_II}. As shown by \citet{Correa_2015}, earlier-forming haloes are expected to be more concentrated, naturally leading to increased substructure survival in FCT. This impact is not expected to affect our results as is discussed in the Appendix \ref{sec:convergence}.

To estimate the impact of different regimes of subhalo mass on our results, we constructed four new mock models based on the original simulations. Incidentaly,  the CDM case  can act as a proxy for a warm DM model (WDM) by removing all substructures with $M_{h} < 10^9\ M_{\odot}$ from the CDM simulation (see \citealt{li_2016}). We then applied the same substructure cut to the $N_b$ and FCT models, with the aim to separate the effect of substructures from this low mass range. Finally, to isolate the contribution of dark haloes, we created an additional model using only substructures with $M_{h} < 10^9\ M_{\odot}$ from the FCT simulation. These configurations are summarised in Table~\ref{tab:substructure_simulation}.

\begin{table}
\caption{Subsample definitions.}    
\label{tab:substructure_simulation}
\centering
\renewcommand{\arraystretch}{1} 
\begin{tabular}{@{}ccc@{}}
    \hline
    Name & Mass Range [$M_{\odot}/h$] & Source Simulation \\
    \hline
    WDM            & $10^9 - 10^{13.6}$ & CDM \\
    NB $>10^9$     & $10^9 - 10^{13.6}$ & NB \\
    FCT $>10^9$    & $10^9 - 10^{13.6}$ & FCT \\
    FCT $<10^9$    & $10^{8.2} - 10^9$  & FCT \\
    \hline
\end{tabular}
\tablefoot{
Names of subsamples from the original simulations. 
Each configuration is defined by the specific substructure mass range selected.
}
\end{table}

Having defined these subsamples, we now analyse their spatial distribution to search for statistical differences.

\begin{figure*}[t]
    \centering
    \includegraphics[width=1\textwidth]{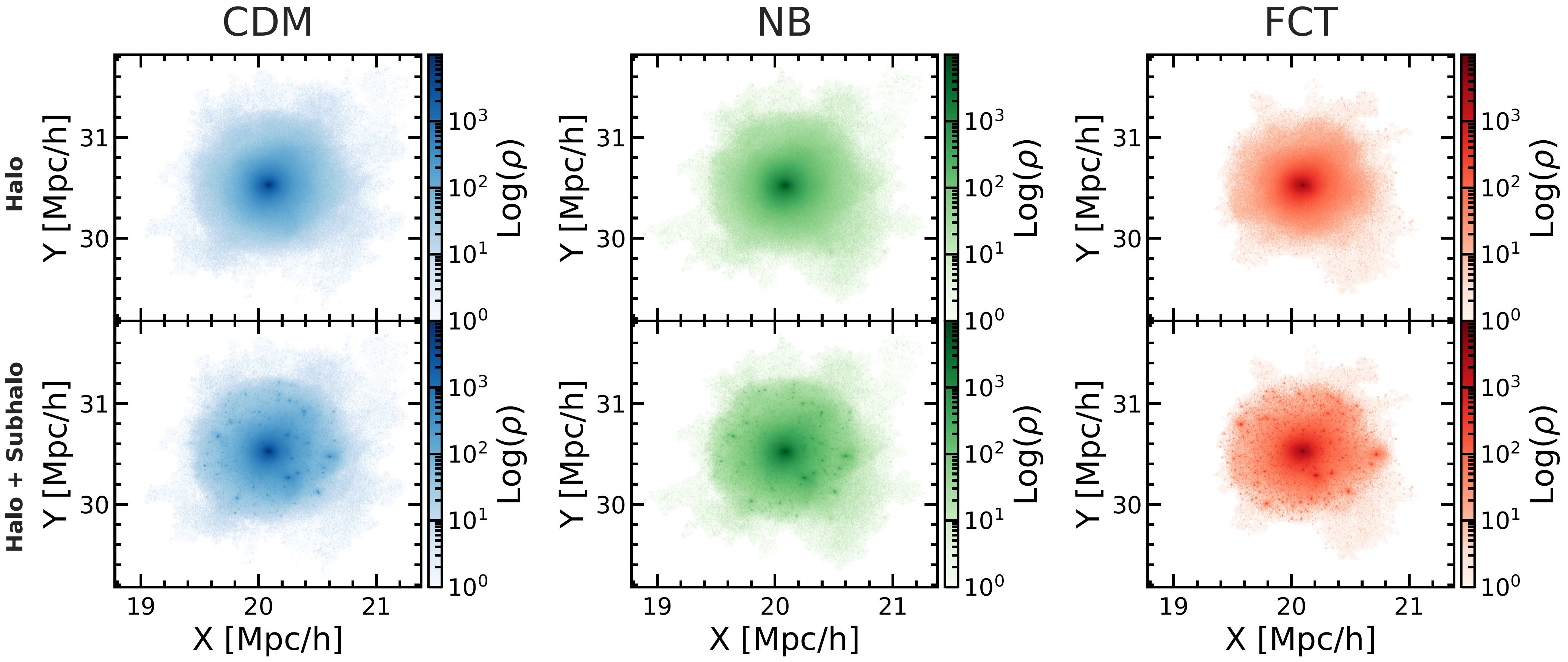}
    \caption{{Projected hexabin distribution of a representative halo at $z=0.5$ in the three simulation models: CDM (left), NB (middle), and FCT (right). Each panel shows a single projection. The top row displays the host halo particles, while the bottom row includes both host and subhalo particles. The inclusion of substructure is visibly more pronounced in the FCT model, highlighting the effect of PBHs on the spatial distribution of matter within haloes.}}
    \label{fig:position}
\end{figure*}

In Figure~\ref{fig:position}, we display a representative matched halo  across different models (different columns), with and without the inclusion of substructures (bottom and top, respectively). We find that substructures in different models show clear differences in their spatial distributions. Our goal is to develop a statistic to quantify these differences.

We analyse the projected density field using 2D hexagonal binning. For each halo, we compute the numerical density $\rho(x,y)$ in hexabins after projecting along three orthogonal directions, yielding three independent measurements per halo.

Before stacking, we centred each halo projection at the rockstar halo centre (halo centre from this point on), rescaled coordinates by $r_{\mathrm{vir}}$, and aligned the principal axes via the 2D inertia tensor:
\begin{equation}
I_{xx} = \frac{1}{N} \sum_{i=1}^N x_i^2, \quad I_{yy} = \frac{1}{N} \sum_{i=1}^N y_i^2, \quad I_{xy} = \frac{1}{N} \sum_{i=1}^N x_i y_i.
\end{equation}
The eigenvectors of $I$ define the rotation matrix for alignment, and the eigenvalues are used to construct elliptical annuli.

Across the three models, the resulting mean ellipticity averaged over all halo projections are indistinguishable within the errors, indicating that at least within this halo mass range the presence of PBHs does not affect the shapes of haloes; still, it should be born in mind that our sample is rather small and differences may lie within the errors.

Within each elliptical annulus, we compute the median density $\tilde{\rho}_{\mathrm{annulus}}$ from the enclosed hexabins. The median is preferred over the mean to represent a typical background value, as it is less sensitive to extreme densities introduced by substructures. Each hexabin was then assigned a local density contrast,
\begin{equation}
\delta(r) = \frac{\rho_{\mathrm{hex}}(r)}{\tilde{\rho}_{\mathrm{annulus}}(r)} - 1,
\end{equation}
that could, in principle, manifest as measurable signals in observations such as brightness fluctuations, although details on the stellar-mass to halo mass relation are important for a quantitative prediction which we defer to future work. It could also be possible with large stacks of strong lenses to input $\delta(r)$ as an ingredient to fit the results.

This yields a normalised $\delta$ field that highlights deviations from the local profile, enhancing substructure detection. The analysis is restricted to elliptical radii in the range $[0.1, 0.85]\ r_{\mathrm{vir}}$ to avoid resolution effects and empty bins near the virial boundary.

 The projected density contrast ($\delta(r)$) allows us to compare the internal profiles of haloes among the different models by normalizing the radial coordinate to the virial radius and averaging over our halo sample.

\begin{figure}[t]
    \centering
    \includegraphics[width=0.5\textwidth]{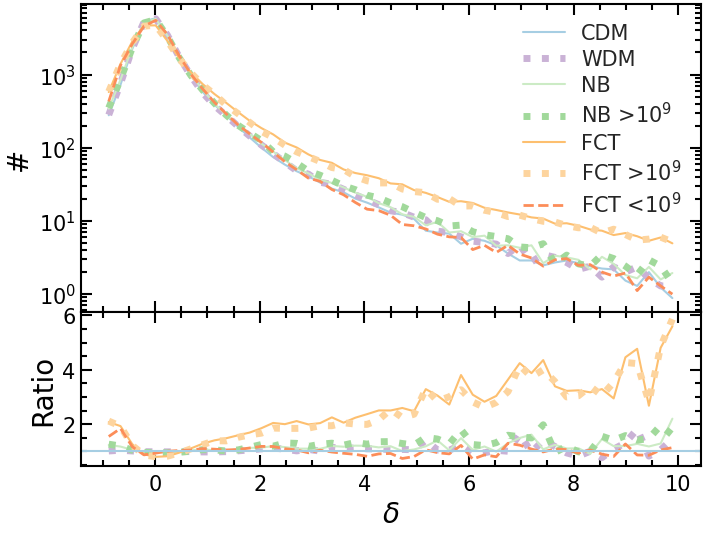}
    \caption{{Stacked $\delta$ distribution for the 17 most massive haloes, each analysed across three independent projections. The upper panel shows the normalised count of $\delta$ values, while the lower panel displays the ratio of each model relative to the CDM baseline. The FCT model, particularly when including only subhaloes with $M_h > 10^9\ M_{\odot}$ (FCT > $10^9$), exhibits a clear excess at high $\delta$ values compared to other models. This result supports the potential for detecting PBH-induced substructure through  brightness fluctuations.}}
    \label{fig:delta.png}
\end{figure}

Figure~\ref{fig:delta.png} shows the average $\delta$ distribution across the three projections of the 17 haloes. The lower panel displays the ratio relative to the CDM model. The FCT model exhibits a clear excess, indicating a measurable difference between scenarios with and without PBHs. In contrast, the NB and WDM models are harder to distinguish with this statistic alone. Most of the FCT signal arises from subhaloes with $M_{h} > 10^{9}\ M_{\odot}$, as seen in the dotted line corresponding to the FCT $>10^{9}$ sample, which may include a luminous component.

While the average $\delta$ profiles provide insight into overall trends, we next study the variance of $\delta$ as a function of radius. This helps capture fluctuations and highlight where the models diverge most significantly.

\begin{figure}[t]
    \centering
    \includegraphics[width=0.5\textwidth]{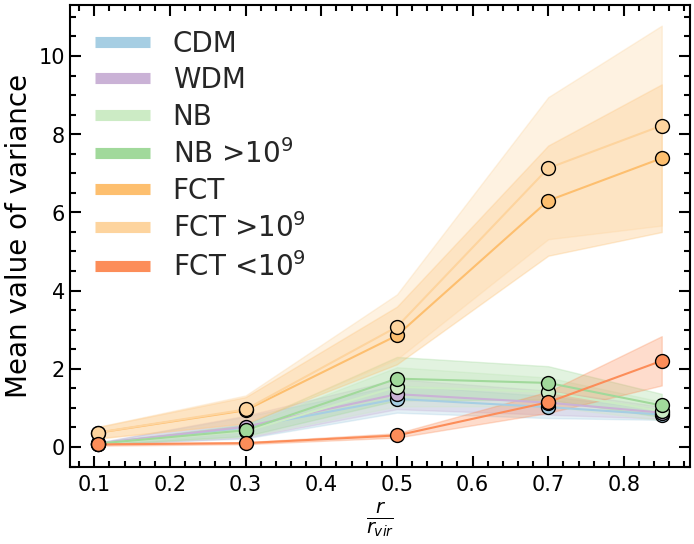}
    \caption{{Mean variance of the $\delta$ distribution measured in five radial bins normalised by the virial radius, $r_{\mathrm{vir}}$, for each model. The FCT model shows the largest variance, especially near $r_{\mathrm{vir}}$, highlighting the effect of PBH-induced substructures. We also explore the separate contributions of haloes with $M_h > 10^9\ M_{\odot}$ (potentially luminous) and those with $M_h < 10^9\ M_{\odot}$ (dark subhaloes). The larger signal from the luminous component suggests that observational detection through radial luminosity fluctuation profiles may be a promising method to constrain PBH scenarios.}}
    \label{fig:varianza.png}
\end{figure}

We compute the variance of $\delta$ for each model as a function of $r/r_{\mathrm{vir}}$, evaluating the dependence of the dispersion of the median with radius. The results are presented in Figure~\ref{fig:varianza.png}. The shaded areas around each curve represent the $1\sigma$ confidence intervals estimated using a bootstrap resampling method. We find that different models can be discriminated at distinct radial scales: while the inner regions of haloes show only mild differences, the outskirts—near $r_{\mathrm{vir}}$—prove most informative for identifying PBH-induced effects in the FCT model.

To complement the previous metrics, we analyse the number density of subhaloes as a function of radius, separated into different mass regimes. This breakdown allows us to isolate the contributions of dark, luminous, and massive substructures to the overall signal, within the context of the mass ranges defined earlier.

\begin{figure*}[t]
    \centering
    \includegraphics[width=1\textwidth]{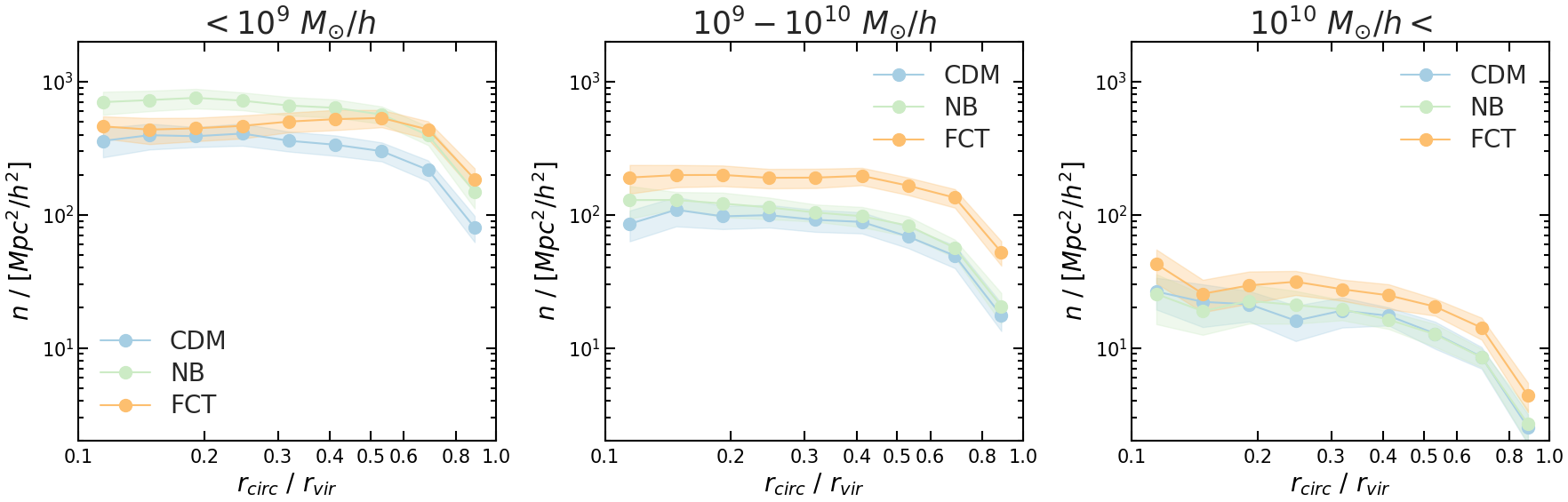}
    \caption{{Normalised numerical density of subhaloes as a function of circularised radial distance ($r_{\mathrm{circ}}/r_{\mathrm{vir}}$) for three substructure mass regimes: $M_h < 10^9\ M_{\odot}$ (left), $10^9 < M_h < 10^{10}\ M_{\odot}$ (centre), and $M_h > 10^{10}\ M_{\odot}$ (right). Results are shown for the CDM, NB, and FCT models, stacked over the 17 most massive host haloes. The shaded areas represent $1\sigma$ bootstrap errors. While NB shows an excess of low mass subhaloes throughout the halo, the FCT model appears to redistribute part of this population toward intermediate and higher masses.}}
    \label{fig:densidad_numerica.png}
\end{figure*}

We are interested in the spatial distribution of three substructure mass regimes: dark haloes ($M_{h} < 10^9\ M_{\odot}$), haloes with a potentially luminous component ($10^9 < M_{h} < 10^{10}\ M_{\odot}$), and the most massive substructures ($M_{h} > 10^{10}\ M_{\odot}$). In Figure~\ref{fig:densidad_numerica.png}, we show the normalised number density of subhaloes as a function of their circularised radial distance, stacked across all host haloes.

By circularised radius we refer to the elliptical transformation:
\begin{equation}
 r_{\mathrm{circ}}^2 = x^2 + y^2 \cdot \left(\frac{a}{b}\right)^2,
\end{equation}
where $a$ and $b$ are the semi-major and semi-minor axes of the halo. This definition allows consistent radial comparisons across projections with varying ellipticity. Here, $x$ denotes the coordinate along the principal axis of the projection (i.e., the direction of elongation), and $y$ the orthogonal axis in the plane of the substructure. Depending on the projection, $x$ and $y$ correspond to different combinations of the original simulation axes (e.g., $x$–$y$, $x$–$z$, or $y$–$z$ planes).

While both the FCT and NB models exhibit an excess of substructure compared to CDM, their spatial distributions differ significantly. The NB model shows a higher abundance of low mass haloes across nearly the entire radial range. In contrast, the FCT model appears to shift part of this low mass population toward intermediate masses, particularly in the outer regions.

This difference suggests that distinguishing between these two scenarios may be possible by analysing the inner regions of galaxy clusters, where the effects of PBHs could even be more pronounced. In the high-mass regime, all models tend to overlap more, although FCT maintains a slight high. These results imply that PBHs may increase both the abundance and mass of substructures, potentially contributing more luminous, low mass subhaloes than models with a blue-tilted spectrum alone, but this increase is less marked for the highest subhalo masses

Furthermore, both NB and FCT models introduce substructure signals strong enough to potentially produce lensing perturbations. \citet{natarajan2024stronglensinggalaxyclusters} review the broader role of galaxy clusters as gravitational lenses to probe DM and cosmology, while \citet{gilman_turbocharging_2024} explore how galaxy–galaxy lensing with JWST can constrain the subhalo mass function. Although our work does not model lensing directly, it provides a key ingredient: an enhanced abundance of subhaloes. For instance, \citet{Natarajan_2017} compared lensing-inferred subhalo abundances in Abell 2744 with $\Lambda$CDM predictions and found a notable discrepancy (see their Figure 10). The abundance ratios presented in our work could be used to quantify such differences, as we find a factor of at least two more substructure in alternative models compared to CDM at $\sim10^9,M_\odot$.

Estimating the impact of our substructure excess on lensing observables would require constructing light cones and applying either semi-analytic lensing models or full hydrodynamic simulations, which we leave for future work.

We show the average values for the full sample of $17 \times 3$ projections, which reveal distinct trends. In the case of the NB model, there is a noticeable excess of dark haloes in the inner regions of the host haloes compared to the other models. On the other hand, the FCT model exhibits a significant excess of haloes with possible luminous components, particularly in the outer regions close to $r_{\mathrm{vir}}$, offering a promising avenue for observational tests. By measuring the brightness fluctuations as a function of distance from the centre of galaxy clusters, it may be possible to distinguish between these models.

Aside from the avenues to explore the abundance of subhaloes explored in this section, of looking for dark subhalo abundances through perturbations of strong lenses, and brightness fluctuations in groups and clusters of galaxies due to the excess of subhaloes in the NB and FCT models, an alternative is to use halo occupation-based methods such as HOD, eHOD, SHAM, or SHAMe \citep{Contreras_2024}.  These can be employed to model the underlying subhalo population.

\section{Conclusions}
\label{sec:conclusions}

We have presented a numerical study of the impact of primordial black holes (PBHs) with an extended mass distribution, characterised by a peak at $M^* = 100 \, M_\odot/h$, assuming a total PBH fraction of $f_{\mathrm{PBH}} = 1$, on the abundance of haloes and subhaloes at redshift $z=0.5$.\footnote{We note that only a small fraction of the total PBH density lies near $M^\ast$ in the extended mass function used} Using cosmological simulations with modified initial power spectra, we compared three scenarios: a standard $\Lambda$CDM model, a model with a blue-tilted primordial spectrum (NB) \citet{Inomata_2017, Clesse_2015}, and a model including both the blue spectrum and a population of PBHs (FCT) with an extended Press–Schechter mass function \citet{Sureda_2021} and including Poisson fluctuations due to the action of unresolved PBHs on the power spectrum we imprint in our initial conditions. (\citealt{li_2016, Padilla_2021}).

The abundance of haloes of high masses remains largely unchanged, consistent with current observational constraints on galaxy group and cluster statistics (\citealt{Abdullah}). We find that PBHs produce a robust and significant excess of low mass haloes in our simulations, particularly in the $10^8$–$10^{10} \, M_\odot$ mass range, with enhancements of at least a factor $\sim 3$ relative to $\Lambda$CDM. This signal is stable across resolutions and not dominated by numerical artefacts, as verified through convergence tests and concentration-based analyses. Even under the most conservative assumptions about numerical disruption, the excess of substructure we measure represents a robust lower limit.

The models begin to show significant differences at intermediate and low masses, below $10^{11}\, M_\odot$. At $z = 0.5$, haloes below $10^{11},M_\odot$ can only be probed indirectly, for example, through the faint end of the galaxy luminosity or stellar mass functions. However, uncertainties in galaxy formation models—including stellar and gas mass estimates, dust production, and colour distributions—introduce additional complications  \citet{Contreras_2024}.

The comparison to ellipsoidal collapse mass functions, in particular obtained via the SMT formalism \cite{ST}, shows departures of varying degrees around $10\%$, indicating the need for numerical simulations to produce accurate predictions of halo abundances in alternative models.

We also looked at the abundance of substructures in massive haloes in the simulations (making note that this analysis and the following ones can only be done with numerical simulations). The PBH-induced excess is most prominent in subhaloes which can be up to a factor of $\sim6$ more abundant with respect to $\Lambda$CDM. This excess is more clearly seen in the outer regions of massive host haloes. These properties suggest that observational strategies targeting strong lensing perturbations or radial luminosity fluctuations in galaxy clusters may be effective in constraining PBH scenarios. While our simulations focus on galaxy-group scales at $z = 0.5$, we expect similar trends to arise in cluster environments. Testing this requires larger-volume simulations including more massive haloes, which we leave for future work.

Our results also show that blue spectral index models without PBHs (NB) can also generate an important excess particularly in the subhalo abundance in massive haloes, which opens the possibility to constrain small scale fluctuation excesses expected from such alternative inflation scenarios along with a conservative lower bound for PBH models which induce additional Poisson effects.

In these massive haloes (comparable to galaxy groups), the amplitude of radial density fluctuations reveals clear model-dependent differences that could be exploited using future strong lensing perturbations \citep{gilman_turbocharging_2024} and brightness fluctuation observations to discriminate between dark-matter scenarios.

These results correspond to FCT extended mass functions for PBHs with a characteristic mass of $M^*=100M_\odot$ that make up all of the DM in the simulation (a possibility according to current constraints; see \citealt{Sureda_2021}). However, the initial power spectrum of this model is within $10\%$ of that corresponding to extended mass functions with a higher mass cut-off at $10^4M_\odot$, with a fraction of DM in PBHs of only $0.5\%$ \citealt{Colazo_II}.  This highlights degeneracies in the PBH parameter space, on the one hand, but it also illustrates the strong effect of even a low abundance of PBHs on the low mass end of the halo and subhalo mass function, on the other.

The enhanced abundance and distinct spatial distribution of substructure in PBH models may offer testable signatures in upcoming surveys such as Roman, Euclid, and JWST. This work offers predictions that may help guide future analyses and contributes to exploring new avenues for constraining the PBH parameter space.

\begin{acknowledgements}
This work used computational resources from UNC Supercómputo (CCAD) – Universidad Nacional de Córdoba (https://supercomputo.unc.edu.ar), which are part of SNCAD, República Argentina. PC acknowledges support from a CONICET Doctoral Fellowship. NP acknowledges support from PICT-2022-00700 and PICT-2023-0002 of FonCyT, Argentina. We are grateful to the anonymous referee for their constructive feedback, which contributed to strengthening this work.  We thank Sebastián Gualpa, Carolina Villalón and Lucas Andrada for their assistance with pipelines and software implementations. We also gratefully acknowledge the support of institute staff members Viviana Bertazzi and Mabel López.

\end{acknowledgements}

\bibliographystyle{aa} 
\bibliography{aa55582-25}

\begin{appendix}
\onecolumn

\section{Estimating possible effects from numerical noise}
\label{sec:convergence}

We explore the possible effects of numerical errors on our results:

(i) Reducing the particle mass by a factor of $\simeq 5.4$ (from the $35\,{\rm Mpc}\,h^{-1}$ box to the $20\,{\rm Mpc}\,h^{-1}$ box) changes the subhalo mass function (SMF) by $\le 0.04$\,dex ($\approx10\%$) over $8.2 \le \log (M/M_\odot) \le 11$ for all cosmologies (Fig.~\ref{fig:ratio_simus}).\footnote{The ratio in Fig.~\ref{fig:ratio_simus} is defined as $\Delta\log_{10}N = \log_{10}N_{\rm 35\,Mpc} - \log_{10}N_{\rm 20\,Mpc}$.} Similar convergence analyses have demonstrated that robust statistical properties (not the individual properties of subhaloes) such as the subhalo mass function can remain stable even below the traditional 100 particle threshold \citep{springel_aquarius_2008,Onions_2012,Knebe_2013,van_den_Bosch_2016,Griffen_2016}.
\citet{Onions_2012} show that mass and maximum circular velocity can even be inferred reliably with particle numbers near of 20. All our data, such as our estimation of $r_s$  using Klypin's estimation (\citealt{Klypin11}, see also \citealt{Behroozi_2012}) to determine the concentration distribution of subhaloes.  

We estimate the Klypin scale radius $R_s$—a robust surrogate for the NFW scale radius in low particle haloes—using the measured virial mass $M_{\mathrm{vir}}$ and maximum circular velocity $v_{\max}$. For an NFW profile the radius at which $v_{\max}$ occurs is a fixed multiple of $R_s$, namely $R_{\max}=2.1626\,R_s$.  Equating the enclosed mass at $R_{\max}$ to the dynamical mass that yields $v_{\max}$ leads to  
\begin{equation}
\frac{c}{f(c)} \;=\; \frac{v_{\max}^{\,2}\,R_{\mathrm{vir}}}{G\,M_{\mathrm{vir}}}\;
\frac{2.1626}{f(2.1626)}, 
\label{eq:klypin_c}
\end{equation}
where $
f(x)=\ln(1+x)\;-\;x/(1+x)$,
and $c\equiv R_{\mathrm{vir}}/R_s$ is the concentration.  Inverting Eq.~(\ref{eq:klypin_c}) numerically yields $c$ and hence $R_s$.  This 'Klypin radius' is preferred over direct profile fitting for haloes with $\lesssim100$ particles, as it is less sensitive to noise and to the limited resolution of the innermost regions. It is possible that simulations with additional small scale fluctuations as FCT make the halo profiles differ somewhat from a NFW shape.  Therefore, using this method could introduce systematic offsets in the concentration values found for the NB and PBH simulations.  Therefore we do not take the concentration values of these latter models as exact but rather as indications of relative increases in typical concentrations of haloes with respect to CDM.

(ii) Increasing the minimum particle threshold from $N_{\rm part}=30$ ($10^{8.2} M_\odot$) to $N_{\rm part}=100$ ($10^{8.72} M_\odot$) changes the FCT/CDM ratio by $<1\%$ (Fig.~\ref{fig:Percentage}). 

Across the intermediate mass range $8.5 \lesssim \log (M/M_\odot) \lesssim 10$, the subhalo mass function in CDM preserves the slope $\mathrm{d}N/\mathrm{d}\log M \propto M^{-0.9}$ at both resolutions. This suggests that global over-merging (see \citet{van_den_Bosch_2018}) is not dominant in our sample, although it cannot be entirely ruled out.

\begin{figure*}
  \centering
  \includegraphics[width=\linewidth]{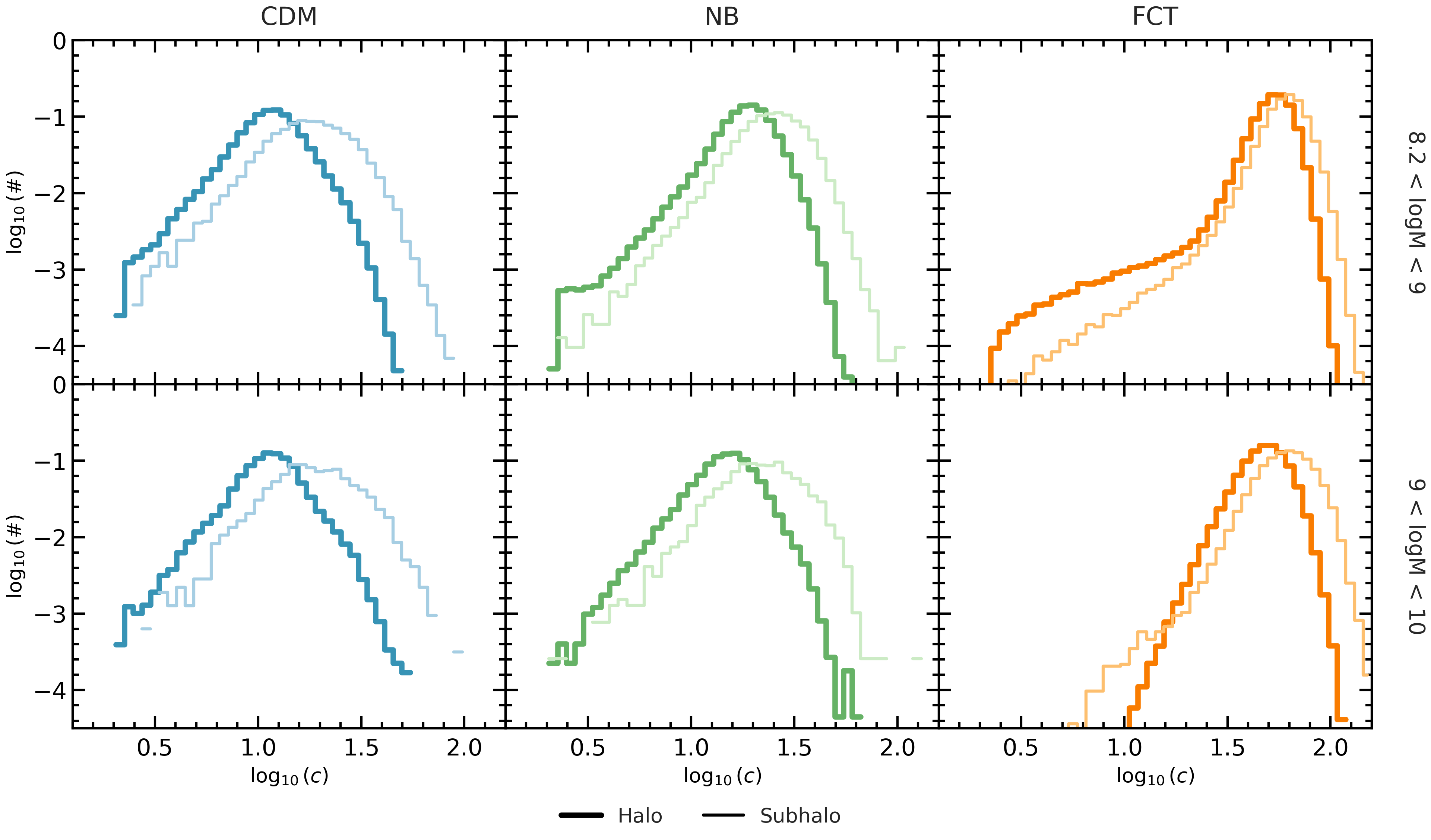}
  \caption{Concentration distributions at $z = 0.5$ for the three cosmologies.  Thick solid lines:  {host} haloes; thin solid lines:  {subhaloes}. Top row: $8.2<\log M/M_\odot<9$; bottom row: $9<\log M/M_\odot<10$. Host–subhost slopes are nearly parallel in every model; CDM displays the smallest separation albeit potentially suffering the greater fragility of its low concentration objects.}
  \label{fig:conc_dist}
\end{figure*}

\begin{figure*}
  \centering
  \includegraphics[width=\linewidth]{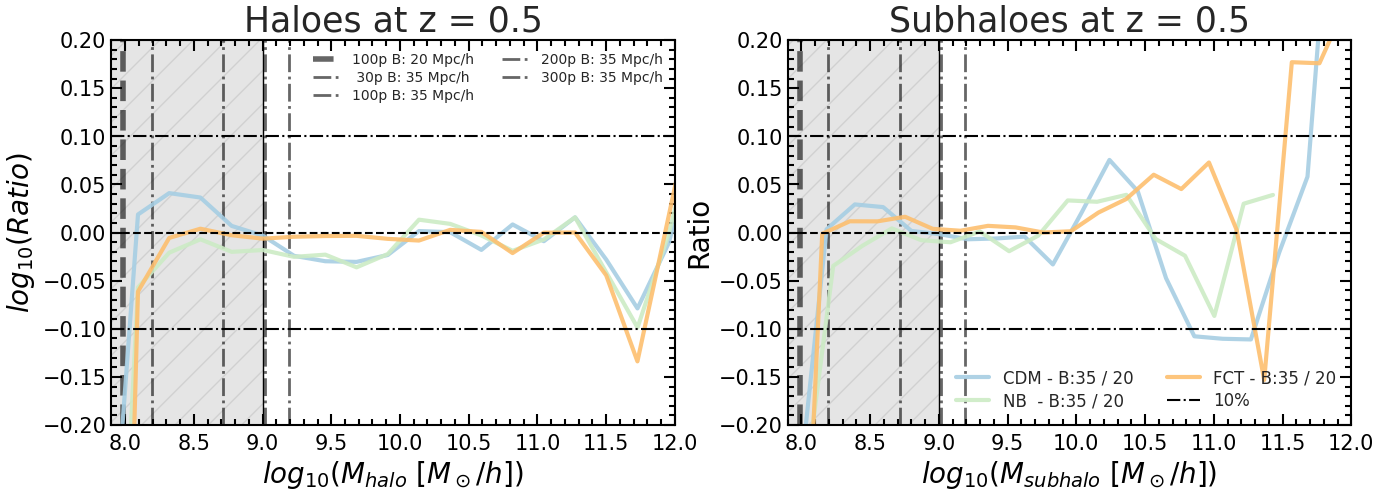}
  \caption{Logarithmic difference between the $35$ and $20$\,Mpc\,$h^{-1}$ boxes, $\Delta\log_{10}N = \log_{10}N_{\rm 35\,Mpc} - \log_{10}N_{\rm 20\,Mpc}$, for haloes (left) and subhaloes (right). Horizontal dash-dotted lines mark $10\%$ differences; the curves stay well within this band except below the cut of $8.2 M_\odot$.}
  \label{fig:ratio_simus}
\end{figure*}

The concentration distribution of host and subhost slopes are nearly parallel in all cosmologies (Fig.~\ref{fig:conc_dist}) (perhaps less so in FCT); CDM shows the smallest offset, as expected if low-$c$ objects are the most fragile. NB and FCT haloes, being systematically more concentrated (bearing in mind that concentration is obtained using Klypin's method in non CDM models), are intrinsically harder to disrupt; a strong numerical loss in CDM would produce a visible deficit in the concentration histogram, which is not observed.

We note that \citet{van_den_Bosch_2018} cautions that even subhaloes resolved with $\sim10^6$ particles can suffer artificial stripping, yet also stresses that discreteness noise affects high- and low mass losses. The final effect discreteness noise does not strongly affect the average mass-loss rate. Because our work focuses on ensemble statistics—the subhalo mass function, radial trends, and concentration histograms—rather than on the properties of individual subhaloes any residual noise should not be dominant. Convergence tests, the particle-number cut, and the physical motivations together confine systematic drifts in the FCT/CDM ratio to $\lesssim10\%$, well below the factor $\gtrsim6$ excess we detect.  Hence our statistical conclusions appear robust.

\end{appendix}

%
%

\end{document}